# Temperature Dependence of the Dynamics of Portevin-Le Chatelier Effect in Al-2.5%Mg alloy

A. Chatterjee[1], K.L. Murty[2], N. Gayathri[1], P. Mukherjee[1] and P. Barat[1*]

[1] *Variable Energy Cyclotron Centre, 1/AF Bidhan Nagar, Kolkata 700064, India*

[2] *North Carolina State University, Nuclear Engineering Department,
2113 Burlington Engineering Laboratories, 2500 Stinson Drive,
Raleigh, NC 27695, USA*

Abstract:

*Tensile tests were carried out by deforming polycrystalline samples of Al-2.5%Mg alloy at four different temperatures in an intermediate strain rate regime of $2 \times 10^{-4} s^{-1}$ to $2 \times 10^{-3} s^{-1}$. The Portevin-Le Chatelier (PLC) effect was observed throughout the strain rate and temperature region. The mean cumulative stress drop magnitude and the mean reloading time exhibit an increasing trend with temperature which is attributed to the enhanced solute diffusion at higher temperature. The observed stress-time series data were analyzed using the nonlinear dynamical methods. From the analyses, we could establish the presence of deterministic chaos in the PLC effect throughout the temperature regime. The dynamics goes to higher dimension at a sufficiently high temperature of 425K but the complexity of the dynamics is not affected by the temperature.*

*Keywords:* Portevin-Le Chatelier effect; Aluminum alloy; Chaos

## I. Introduction

Plastic flow of dilute metallic alloys may display repetitive yielding within certain ranges of the deformation conditions which under constant applied strain rate is manifested as the serrations in the stress-strain curve. This irregular flow is associated with inhomogeneous strain localization within the material and is commonly referred to as the Portevin-Le Chatelier (PLC) effect [1,2]. In recent years, the study of the PLC effect has drawn continuing interest since it is a fascinating example of complex spatio-temporal dynamics arising from the collective behavior of various defects within the metallic alloy system. In microscopic sense, the physical origin of the PLC effect is associated with dynamic strain ageing (DSA) due to the interaction of gliding dislocations and diffusing solute atoms [3-7]. The repeated pinning and unpinning of the mobile dislocations by the solute cloud is manifested as the serrations in the stress-strain curve. During the PLC effect, a bunch of dislocations move coherently forming the deformation band. In polycrystals three generic types of deformation bands were identified based on metallurgical taxonomy, namely type A, type B and type C. The deformation band character changes from type C to type B and finally to type A with increasing strain rate and decreasing temperature. Hence, both the strain rate and temperature are very important control parameters in the PLC dynamics. In our previous works, we have carried out tensile tests on Al-2.5%Mg alloy over a wide range of strain rates at room temperature to characterize the PLC dynamics at different strain rates [8-10]. Several works have been performed to study the effect of temperature on the dynamics of the PLC effect. Many investigators reported the changes in the mechanical properties such as yield stress, ductility, critical strain etc. with temperature in different Aluminum alloy systems [11-14]. Variation of strain rate sensitivity with temperature is also reported [15]. But still there is an ample scope to explore the complex dynamics of the PLC effect and to study the effect of temperature on it.

---

[*] Corresponding author Email: pbarat@vecc.gov.in
Ph No.: +91 33 23182401; Fax: +91 33 23346871

The dynamics of the PLC effect is a rare example which exhibits both chaos and self organized criticality (SOC) under different experimental conditions. The underlying dynamics of the PLC effect changes its character from chaos to SOC with increasing strain rate or decreasing temperature [16-19]. No study has been done so far to investigate the effect of temperature on the chaotic dynamics of the PLC effect for a particular strain rate experiment. In this paper, an attempt has been made to understand the influence of temperature on the dynamics as well as the statistical parameters of the PLC effect by extensive analysis of the observed experimental time series data.

## II. Experimental

The Al-2.5% Mg sheet was prepared from ingot by hot rolling process at $450^0$C followed by cold rolling and an intermediate annealing was carried out just before the final stage cold rolling. Tensile tests were conducted on flat specimens prepared from polycrystalline Al-2.5%Mg alloy in a strain rate regime from $2 \times 10^{-4} s^{-1}$ to $2 \times 10^{-3} s^{-1}$. Since we are interested to work in the chaotic regime of the PLC effect which is associated with the type B band, an intermediate strain rate regime has been selected to get PLC serrations of type B during the plastic deformation of the material. Specimens with gauge length, width and thickness of 35, 6.5 and 1.5 mm respectively were tested at four different temperatures (293K, 323K, 373K and 423K) in an INSTRON (model 5566) machine. An external cylindrical furnace is used to carry out the tests at elevated temperatures. Atodiresei et al. [20] reported that in the Al-Mg alloy the precipitates which act as the strong pinning points to the deformation band movement, gradually dissolve at a temperature of 450K or higher. In order to obtain a pronounced PLC effect, we have restricted our studies up to a temperature of 423K. The PLC effect was observed throughout the strain rate and temperature regime. The stress-time response was recorded at a periodic time interval of 0.03s. To extract different dynamical features of the PLC effect, the stress-strain data are analyzed following the methods described in the subsequent sections. Here the stress and strain correspond to the true stress and true strain respectively.

## III. Method of Analysis

The experimental stress-time data exhibit an increasing trend due to strain hardening and the serrations in the stress-time curve are extracted by eliminating the strain hardening part. A typical segment of the stress serrations for the tensile test carried out at a strain rate of $8 \times 10^{-4} s^{-1}$ and temperature 323K is shown in the Fig.1.

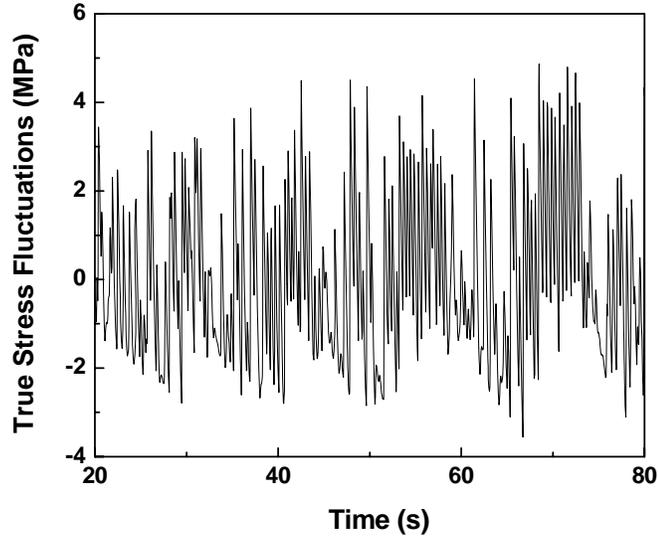

Fig. 1: The segment of the drift corrected true stress fluctuation vs. time curve for the strain rate $8\times10^{-4}$ s$^{-1}$ and temperature 323K.

With increasing temperature, PLC band changes its character from type A to B and successively to type C for a given strain rate experiment. Before proceeding further, it is necessary to ascertain a priory that the observed PLC serrations are of type B only throughout the strain rate and temperature regime studied. It is well known that the stress drop distribution of the type C band is Gaussian in nature whereas that of the type B band is an asymmetric Gaussian [21]. For the serrations of type C, the reloading parts between two successive instabilities are partially plastic [22]. A typical stress drop magnitude distribution is shown in Fig. 2 for the strain rate of $2\times10^{-4}$s$^{-1}$ and temperature 423K which is asymmetric Gaussian in nature. The inset of Fig. 2 represents a portion of the serrated stress-strain curve for the strain rate of $2\times10^{-4}$s$^{-1}$ and temperature 423K which does not exhibit any plastic reloading part. The same behavior of the stress drops are obtained throughout the studied strain rate and temperature range. Hence we can conclude that the serrations obtained in the stress-strain curve are of type B in nature. Since the stress drops originate from the dislocation-solute interaction, all the statistical and dynamical analyses are confined to the serrated part of the stress-time data only.

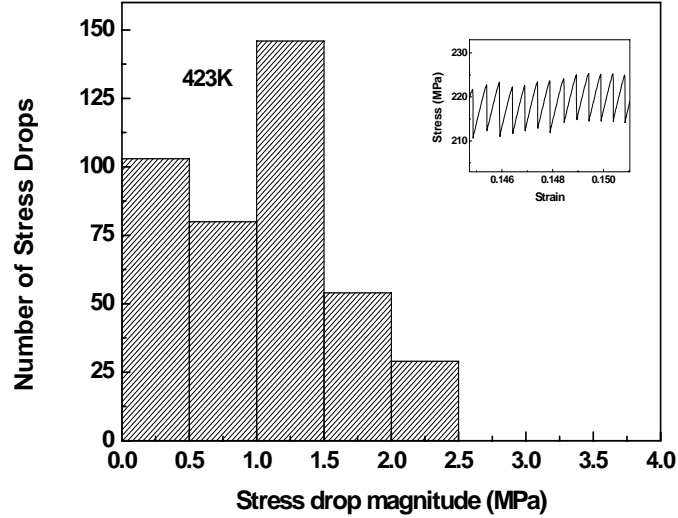

Fig.2: The stress drop magnitude distribution for the experiment carried out at a strain rate of $2\times10^{-4}$ s$^{-1}$ and temperature 423K.

We have investigated the variation of cumulative mean stress drop magnitude $(\Delta\sigma)$ with strain at different test temperatures. The time, during which the dislocation band waits at the obstacle, i.e. the time during which the stress level rises, is designated as the reloading time and the distribution of the reloading time is plotted at different temperatures for a fixed strain rate.

Apart from the statistical analyses, we have carried out few nonlinear time series analyses on the stress-time data to explore various features of the underlying dynamics. PLC effect is a paradigm of complex metallurgical phenomenon which must be associated with a large number of degrees of freedom. Since we are particularly interested in the chaotic regime of the PLC effect, we start our analysis to investigate the presence of chaos in the stress-time series data at different temperatures. S.J. Noronha et al. [23] have established the presence of chaos in Al-Mg alloy by analyzing the stress-time series data. The primary step of the dynamical analysis of the PLC effect is the reconstruction of the multidimensional phase portrait from the one dimensional time series data $\{x_i, i= 1, 2, ..., N\}$ on the basis of the embedding theorems [24,25]. The phase space trajectory in the embedding space is reconstructed with the time delay vectors $X_i = \{x_i, x_{i+\tau}, x_{i+2\tau}, ....., x_{i+(m-1)\tau}\}$, where $m$ represents the embedding dimension and $\tau$ is the delay time. The reconstructed attractor retains all the properties of the original one provided the values of m and $\tau$ are selected properly. The value of *m* is estimated by the false nearest neighbor method [26]. $\tau$ is estimated from the first zero crossing of the auto-correlation function [26]. The random noise present in the experimental stress-time series data is removed by the method of Singular Value Decomposition (SVD) [27]. The sensitivity of a chaotic system on its initial condition is reflected in the presence of at least one positive Lyapunov exponent. Hence, the existence of a positive Lyapunov exponent is one of the most reliable signatures of chaotic dynamics [28]. The estimation of the largest Lyapunov exponent is performed with the method proposed by Wolf et al. [29]. The Grassberger-Procaccia algorithm [30] is used to determine the correlation dimension ν of the attractor. The correlation integral *C(r)* is defined as the fraction of pair of points [*X(i), X(j)*] with separation less than r in the reconstructed space. In mathematical form C(r) is expressed as:

$$C(r) = \lim_{N \to \infty} \frac{2}{N(N-1)} \sum_{i \neq j}^{N} \Theta\left[r - |X(i) - X(j)|\right]$$

*where N is the total number of time delay vectors and Θ(x) is the Heaviside step function with Θ(x) = 1 for x>0 and Θ(x) = 0 for x<0. C(r) exhibits a scaling behaviour $C(r) = r^{\nu}$ for strange attractor.* Hence, ν can be calculated from the slope of the log*C(r)* vs. log*r* plot. For a complex dynamics like that of the PLC effect, it is always interesting to quantify the complexity of the dynamics and investigate if it is having some influence of temperature on it. To compare the complexity of the PLC effect in the substitutional Al-Mg alloy, the Multiscale Entropy (MSE) analysis [31] was carried out on the experimental stress-time data in the PLC regime at different temperatures.

### IV. Results and Discussions

A typical plot of the variation of the cumulative mean stress drop magnitude $(\Delta \sigma)$ with increasing strain at different temperatures is shown in Fig. 3 for the experimental data obtained at a strain rate of $8 \times 10^{-4} s^{-1}$.

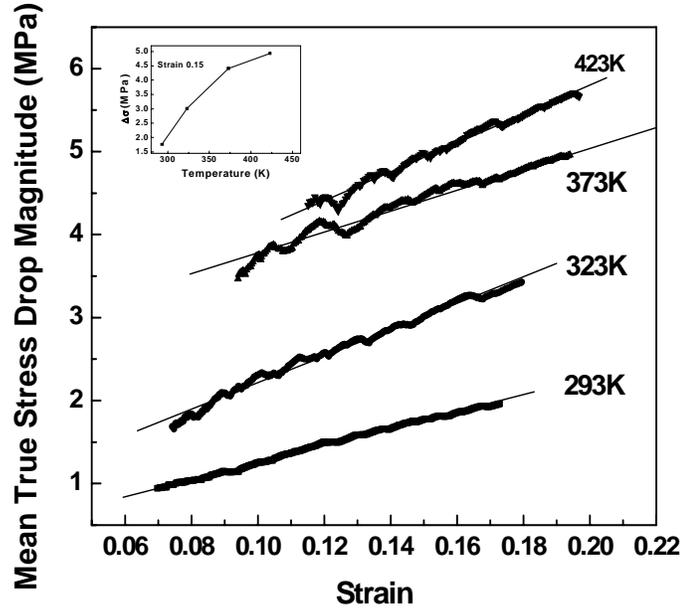

Fig.3: Variation of cumulative mean stress drop magnitude $\Delta\sigma$ with strain at different temperatures for the strain rate of $8\times10^{-4}$ s$^{-1}$. The inset shows the value of $\Delta\sigma$ for different temperatures at a strain value of 0.15.

It is observed that $\Delta\sigma$ increases with strain for any given temperature which is a very common feature of the PLC effect [32,33]. For a particular strain, $\Delta\sigma$ has an increasing trend with temperature as well, as shown in the inset of the Fig. 3. A typical plot of the reloading time distribution at the strain rate of $8\times10^{-4}$s$^{-1}$ for different temperatures is presented in Fig. 4. The distribution shows a shift of its mean to the higher value at elevated temperatures. The inset of the Fig. 4 shows the variation of mean reloading time with temperature. All these observations can be explained in terms of the enhanced solute diffusion with increasing temperature. Solute atoms diffuse towards the dislocations under the stress fields around its core. For substitutional Al-Mg alloy, several mechanisms were proposed for the diffusion of Mg in Al. Among them few are vacancy assisted, such as the nearest neighbour diffusion and the ring mechanisms [34]. Since the activation energy for the nearest neighbour diffusion (1.228 eV) is lower than that of ring mechanism (1.418 eV), the former one is more favorable mechanism from the energy point of view. There is also a possibility for Mg atom to diffuse in absence of vacancies with the aid of a different kind of ring mechanism where its neighborhood Al atoms are involved in the diffusion process but the corresponding activation energy (3.3 eV) is quite high for it to realize. Hence, it can be concluded that in Al-Mg, the Mg diffusion is vacancy assisted. With increasing temperature the intrinsic vacancy concentration in the material increases which follows from thermodynamic relation as $C_T = C_0 \, e^{-\frac{E_v}{kT}}$, where $C_0$ is the initial atomic density and $E_v$ is the vacancy formation energy. But in Al and its alloys, these thermal vacancies are dominating only at a temperature of 500K or higher [35]. At a temperature regime below 500K, the solute diffusion is controlled by deformation induced vacancies. With increasing strain, more vacancies are created following a relation $C_v^\varepsilon = A\varepsilon^m$ [36-38], where A is a constant, $C_v^\varepsilon$ is the deformation induced vacancy concentration and $\varepsilon$ is the strain; the test temperature range is low enough that the in-situ vacancy annihilation is but a small perturbation [38]. The increased vacancy concentration enhances the solute diffusion. The population of solute atoms locking the band movement increases with strain resulting in a higher pinning barrier. Hence, the deformation band has to spend longer time at the obstacle and the width of the band increases with the aid of dislocation multiplication. When this band tears of the solute cloud, the large band width is manifested in huge stress drop which is reflected in the increase in $\Delta\sigma$ with strain. It must be noted at this point that, even if there are enough vacancies in the material, the solute cannot jump into it unless it has the sufficient vibrational energy to overcome certain energy barrier $E_0$. The probability that a solute can exchange its place with the neighbourhood vacancy is given by jump frequency $\vartheta$, where $\vartheta \sim e^{-\frac{E_0}{kT}}$, which suggests that the jump frequency increases with increasing temperature. This explains the fact that the solute diffusion becomes faster with increasing temperature. Moreover, Nortmann et al. [5,39] and Hahner [40] suggested that at elevated temperatures several types of pipe diffusion along the core of the partials, along the stacking fault ribbon etc. become active which are also responsible for solute to diffuse faster at high temperatures. Hence, at elevated temperatures, the faster solute diffusion draws a huge solute cloud towards the dislocation band and a much higher pinning barrier is achieved than that at lower temperatures. As a consequence, the band has to wait for a longer time at the obstacle which is reflected in the increase in mean reloading time and the mean stress drop magnitude with temperature. Apart from the increase in mean reloading time, the reloading time distribution becomes broader with increasing temperature which is evident from the increasing trend of the standard deviation of the distribution with temperature as shown in the inset of Fig. 4. With increasing temperature, the thermal fluctuations also increase which introduces more variability in the magnitude of the reloading time and results in a broader distribution at elevated temperatures.

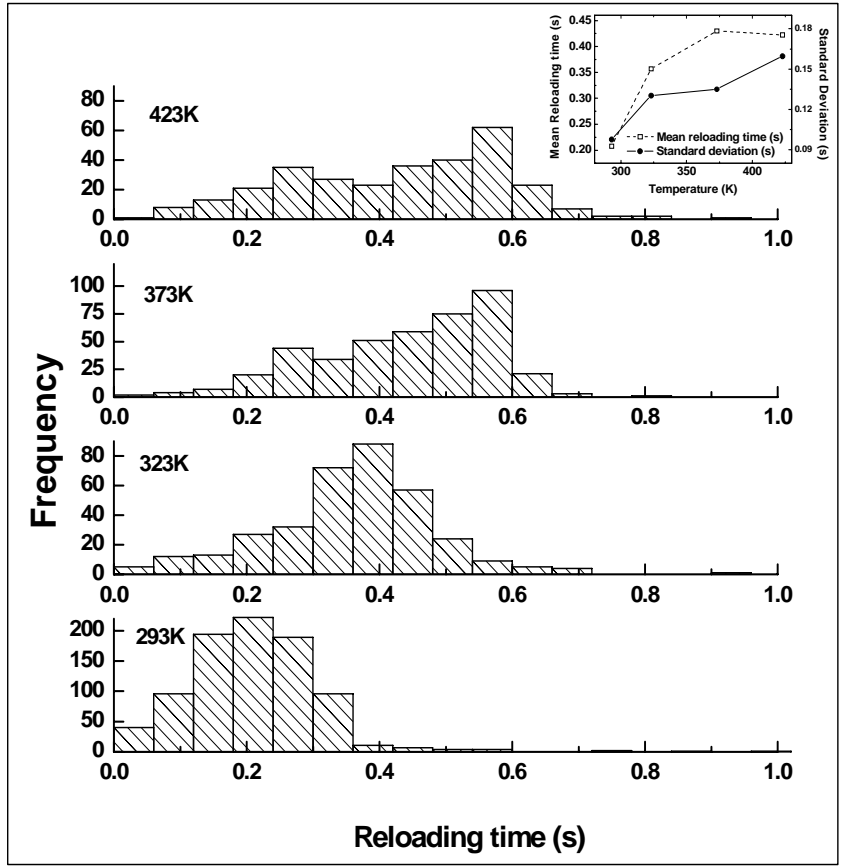

Fig.4: Reloading time distribution at different temperatures for the strain rate of $8\times10^{-4}$ s$^{-1}$. The inset shows the variation of mean reloading time and its standard deviation with temperature for the same strain rate experiment.

Besides investigating the variation in different statistical parameters of the PLC effect like the mean stress drop magnitude and the reloading time with temperature, it is interesting to study the effect of temperature on the underlying dynamics of the PLC effect. To reconstruct the attractor from the time series data, the embedding dimension m is estimated by the false nearest neighbour method to be ~ 10 (~ 15 at 423K) and $\tau \sim 4$ for the stress-time data obtained at all strain rates and temperatures. With these values of m and $\tau$, the phase portrait is reconstructed for each temperature and strain rate. Fig. 5 shows the two dimensional projection of a typical reconstructed attractor for the strain rate of $8\times10^{-4} s^{-1}$ at the temperature of 323K. Here $C_1$ and $C_2$ are eigenvectors corresponding to the two largest eigenvalues. Hence, Fig.5 represents the projection of the reconstructed attractor onto two-dimensional subspace spanned by the first two eigen vectors $C_1$ and $C_2$.

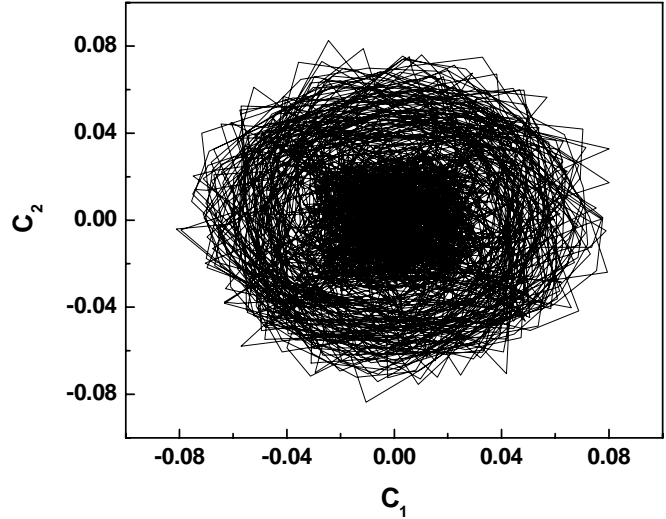

Fig.5: Projection of the reconstructed attractor onto two-dimensional subspace spanned by the first two Eigen vectors $C_1$ and $C_2$ for the strain rate $8\times10^{-4}$ s$^{-1}$ and temperature 323K.

The largest Lyapunov exponent $(\lambda)$ is calculated for the data set of each temperature and strain rate and it appears to be positive in each case, varying between $\lambda \sim 0.6 - 1.0$. This proves the presence of chaos in the underlying dynamics of the PLC effect within this definite strain rate and temperature range. The correlation integral C(r) is calculated by varying the embedding dimension m from 5 to 35. The correlation dimension $\upsilon$ is determined from the slope of the curve logC(r) vs. logr. The slope initially increases with the value of m and saturates after some high value of m is reached. The convergent value of the slope so obtained represents the correlation dimension of the dynamics. Following this formalism, we obtain the value of $\upsilon \sim 3.5$ for room temperature. No appreciable change in the value of $\upsilon$ is obtained until a temperature of 423K is reached. Fig. 6 shows a typical plot of logC(r) vs. logr for the experiment carried out at a strain rate of $8\times10^{-4}s^{-1}$ and temperature 423K by varying the embedding dimension m from 20 to 35 and taking $\tau$ =4.

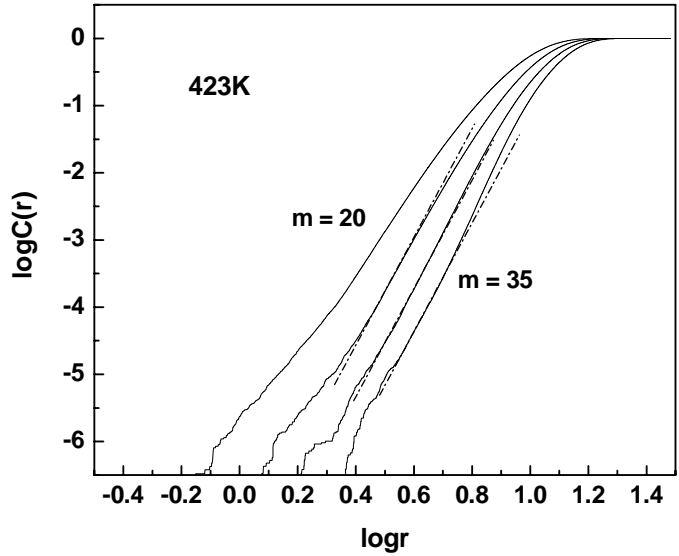

Fig.6: Variation of the Correlation integral $C(r)$ with $r$ with increasing embedding dimension m (20, 25, 30, 35) for the strain rate $8\times10^{-4}$ s$^{-1}$ and temperature 423K.

At this temperature a sudden jump in the value of $\upsilon$ to 7.8 is obtained as shown in Fig. 7. The correlation dimension provides the information about the minimum number of dynamical variables necessary to represent the dynamics of the respective system [41].

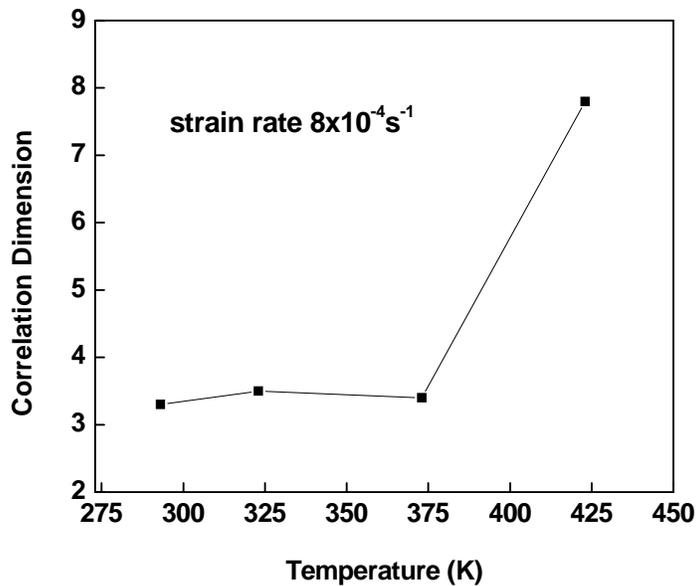

Fig.7: Variation of the Correlation Dimension with temperature for the strain rate of $8\times10^{-4}$ s$^{-1}$.

Hence, it can be concluded that within the strain rate and temperature regime, where the PLC effect exhibits chaotic dynamics, the degrees of freedom exhibits a sudden increase in value at a temperature of 423K. It must be mentioned here that at room temperature the PLC effect in Al-2.5%Mg alloy exhibits SOC for the strain rate ~ $10^{-3}s^{-1}$ or higher which is characterized by power law behavior of the stress drops, the absence of any finite correlation dimension etc. and is dealt in detail else where [18]. In our experiments, we could also identify the presence of SOC in room temperature for the high strain rate regime. Since at all other temperatures, chaotic dynamics were observed through out the strain rate region, we have confined our whole analysis to investigate the effect of temperature on the PLC dynamics in this chaotic regime only.

The sudden increase in the degrees of freedom of the PLC dynamics at high temperature can be explained in terms of the active slip systems available for the propagation of the PLC bands. Since the crystal structure of the Al-Mg alloy is face-centered cubic (FCC), slip occurs in the close packed plane {111} and close packed direction <110> respectively. But slip in FCC material is also observed in several non-octahedral planes. The slip on {110} plane takes place only after a few percent of deformation when the stress level reaches a critical value $\sigma_{110}$ [42]. The value of this critical stress is thermally activated and decreases with increasing temperature. Hence, at elevated temperature, a much lower value of $\sigma_{110}$ is needed which can be achieved during the plastic deformation of the specimen to initiate the slip on {110} planes. Several investigators reported on the existence of the non-octahedral slip planes in Al single crystals as well as polycrystals [42-45]. Hazif et al. [43] reported that at around 443K, there is a transition of the slip planes from {111} to {110} for Al single crystals. Crumbach et al. [45] have shown that for polycrystalline Aluminum alloys the {110} <110> slip is active at high temperatures and the onset of {110} slip was reported near 439K which is very close to the temperature (423K) where the jump in the value of correlation dimension is observed. In this scenario, we anticipate that during the deformation of Al-Mg sample at 423K, additional non-octahedral slip systems have been activated and the dislocations on those planes also participate in the PLC dynamics which is manifested as the increased degrees of freedom of the PLC dynamics at elevated temperature.

In order to quantify the complexity of the PLC dynamics, MSE analysis was carried out on the time series data. A typical result of the MSE analysis is shown in Fig. 8 for the strain rate of $8\times10^{-4} s^{-1}$. It is seen that the Sample Entropy of the PLC data for Al-2.5%Mg alloy does not vary much with temperature for all the scales. This finding leads to the fact that the temperature does not have any influence on the complexity of the PLC dynamics.

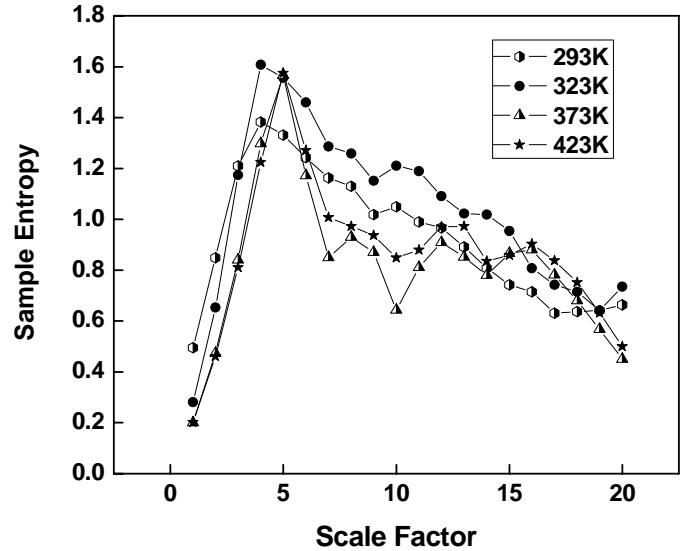

Fig.8: The MSE analysis of the PLC effect in the Al-2.5%Mg alloy at different temperatures deformed at a strain rate $8\times10^{-4} s^{-1}$.

At this point, we should mention one of our earlier works [8] where a comparative study of the PLC effect was carried out between Al-Mg alloy and low carbon steel. The system dimension as well as the complexity was found to be higher in case of low carbon steel than that in Al-Mg alloy in the PLC regime. On the other hand, we observe an increase in dimension of the PLC dynamics in Al-2.5%Mg alloy at a temperature of around 423K but no noticeable change in complexity was observed. The higher dimensionality in the PLC effect in low carbon steel is attributed to the participation of both edge and screw dislocations in DSA. Since screw dislocation is able to cross slip in between the slip planes, collectively the dynamics become more complex. In Al-Mg alloy, at sufficiently high temperature, the hike in the value of dimensionality comes from the activation of new slip systems but here only the edge dislocations are participating in DSA [46,47] which are unable to cross slip. The activation of new slip planes facilitates the edge dislocations on those planes to participate into the PLC dynamics which in turn increases the dimensionality of the system. The inability of those dislocations to cross slip prevents them to change their glide planes during deformation and inhibits the interaction among the edge dislocations from two different slip planes. Since complexity is the measure of how the interwoven parts of a system interact to form the behavior of the whole system, the activation of new slip systems is unable to make the dynamics a more complex one and hence the complexity does not change with temperature.

## V. Conclusions

In conclusion, we have shown that at elevated temperature enhanced solute diffusion results in larger stress drop and increase in reloading time. At sufficiently high temperature activation of new slip systems induces an increase in the degrees of freedom of the dynamics. No effect of temperature on the complexity of the dynamics was observed. In this paper we characterized the PLC effect in the strain rate regime where the underlying dynamics exhibits chaotic nature i.e. the dynamics has a finite dimension. More detailed analysis to investigate the effect of temperature on the PLC effect needs to be done by incorporating wider strain rate regime and extending our analysis to study if the SOC dynamics of the PLC effect also shows some remarkable variation with temperature.

**Acknowledgments**

The experimental work described here was performed in the Nuclear Materials Laboratory at NC State University and we acknowledge the support of the National Science Foundation grant DMR0412583. The assistance of Mr. Usama Elkazaz in carrying out the experiments is acknowledged.